\documentclass[twocolumn,showpacs,prd]{revtex4}
\usepackage{mathrsfs,bm,multirow}
\usepackage{longtable,lscape}
\usepackage{txfonts}
\usepackage{amssymb}
\usepackage{indentfirst}
\usepackage{graphicx,,booktabs}
\usepackage{color}
\usepackage{amssymb}

\begin{document}

\title{Novel charged charmoniumlike structures in the hidden-charm dipion decays of $Y(4360)$}
\author{Dian-Yong Chen$^{1,3}$}
\email{chendy@impcas.ac.cn}
\author{Xiang Liu$^{1,2}$\footnote{Corresponding author}}\email{xiangliu@lzu.edu.cn}
\author{Takayuki Matsuki$^4$}
\email{matsuki@tokyo-kasei.ac.jp}
\affiliation{$^1$Research Center for Hadron and CSR Physics,
Lanzhou University $\&$ Institute of Modern Physics of CAS,
Lanzhou 730000, China\\
$^2$School of Physical Science and Technology, Lanzhou University,
Lanzhou 730000, China\\
$^3$Nuclear Theory Group, Institute of Modern Physics, Chinese
Academy of Sciences, Lanzhou 730000, China\\
$^4$Tokyo Kasei University, 1-18-1 Kaga, Itabashi, Tokyo 173-8602,
Japan}

\begin{abstract}
Studying the hidden-charm dipion decays of the charmoniumlike state
$Y(4360)$, we show that there exist charged charmoniumlike
structures near $D\bar{D}^*$ and $D^*\bar{D}^*$ thresholds in the
$J/\psi\pi^+$, $\psi(2S)\pi^+$ and $h_c(1P)\pi^+$ invariant mass
spectra of the corresponding hidden-charm dipion decays of $Y(4360)$
using the {\it{Initial Single Pion Emission} mechanism}. We suggest
to do further experimental study on these predicted charged
enhancement structures, especially by BESIII, Belle, and the
forthcoming BelleII.

\end{abstract}
 \maketitle

\section{Introduction}
Very recently, the BESIII
Collaboration reported a charged charmoniumlike structure by
studying the $J/\psi \pi^\pm$ invariant mass spectra in $e^+e^-\to
J/\psi\pi^+\pi^-$ at $\sqrt{s}=4.26$ GeV \cite{Ablikim:2013mio}.
This newly observed charged structure around 3.9 GeV is referred to
$Z_c(3900)$, which was confirmed by the Belle Collaboration
\cite{Liu:2013dau}. Later, Xiao {\it et al.} also announced the
observation of $Z_c(3900)$ with a 6$\sigma$ significance by
analyzing 586 pb$^{-1}$ data collected with the CLEO-c detector at
$\psi(4160)$ \cite{Xiao:2013iha}. As indicated in Ref.
\cite{Ablikim:2013mio}, this charged enhancement structure near the
$D\bar{D}^*$ threshold was predicted in Refs.
\cite{Chen:2011xk,Sun:2011uh,Ali:2011ug} before the observation of $Z_c(3900)$.
Among these pioneering theoretical investigations, the {\it{Initial
Single Pion Emission}} (ISPE) mechanism was applied to study the
hidden-charm dipion decays of higher charmonia, where there exist
the charged charmoniumlike structures near the $D\bar{D}^*$ and
$D^*\bar{D}^*$ thresholds in the distributions of the
$J/\psi\pi^\pm$, $\psi(2S) \pi^\pm$ and $h_c(1P)\pi^\pm$ invariant
mass spectra \cite{Chen:2011xk}.

The ISPE is a decay mechanism peculiar to the hidden-charm dipion decays of higher charmonia \cite{Chen:2011xk} and the hidden-bottom dipion decays of higher bottomonia \cite{Chen:2011pv,Chen:2011pu}. The ISPE mechanism was first proposed to explain why the charged bottomoniumlike structures $Z_b(10610)$ and $Z_b(10650)$ exist in the $\Upsilon(nS)\pi^\pm$ $(n=1,2,3)$ and $h_b(mP)\pi^\pm$ ($m=1,2$) invariant mass spectra of the hidden-bottom decays of $Y(10860)$, which was reported by Belle \cite{Collaboration:2011gja}.
The authors adopted the ISPE mechanism further to study hidden-bottom decays of $Y(11020)$ \cite{Chen:2011pu}, and predicted the charged bottomoniumlike enhancement structures similar to $Z_b(10610)$ and $Z_b(10650)$. Recently, we have developed the ISPE mechanism to propose the {\it Initial Single Chiral Particle Emission} (ISChE) mechanism, where the charged charmoniumlike structures with the hidden-charm and open-strange channel were predicted \cite{Chen:2013wca}.

Although we have already presented abundant phenomena of charged charmoniumlike/bottomoniumlike structures
in the hidden-charm/hidden-bottom dipion decays of higher charmonia/bottomonia,
it is not the end of the story. The fact that BESIII confirmed our prediction \cite{Ablikim:2013mio} again
provokes our interest into predicting more novel charged charmoniumlike structures in other processes, which can be accessible at future experiments, especially at BESIII, Belle, and the forthcoming BelleII.

In this work, we notice the similarity between properties of $Y(4360)$ and $Y(4260)$. As the first charmoniumlike structure reported in the $e^+e^-\to J/\psi\pi^+\pi^-$ process, $Y(4260)$ was observed by BaBar \cite{Aubert:2005rm}. Later, both the CLEO and Belle Collaborations
confirmed $Y(4260)$ in the $e^+e^-\to J/\psi\pi^+\pi^-$ process
\cite{Coan:2006rv,He:2006kg,:2007sj}.
$Y(4360)$ was reported by BaBar \cite{Aubert:2006ge,Lees:2012pv} after
analyzing the $\psi(2S) \pi^+\pi^-$ invariant mass spectrum
of $e^+e^-\to \psi(2S) \pi^+\pi^-$, which was confirmed by Belle
\cite{:2007ea}. At present, $Z_c(3900)$ was observed in the $J/\psi\pi^\pm$ invariant mass spectrum of $Y(4260)\to J/\psi\pi^+\pi^-$.
The similarity between $Y(4360)$ and $Y(4260)$ just mentioned above provides us interest in studying whether a similar charged charmoniumlike structure can be found in $Y(4360)\to J/\psi\pi^+\pi^-$ and other hidden-charm dipion decays of $Y(4360)$. To answer to this question, we need to carry out a more in-depth study of the hidden-charm dipion decays of $Y(4360)$.

This paper is organized as follows. After the introduction, we present the hidden-charm dipion decays of $Y(4360)$ discussed above and the calculation of the corresponding decay amplitudes and the differential decay widths. In Sec. \ref{sec3}, the numerical results are given. The last section devoted into a short summary.

\section{The hidden-charm dipion decays $Y(4360)$}\label{sec2}

\begin{figure}[htp]
\begin{tabular}{cc}
\includegraphics[width=0.22\textwidth]{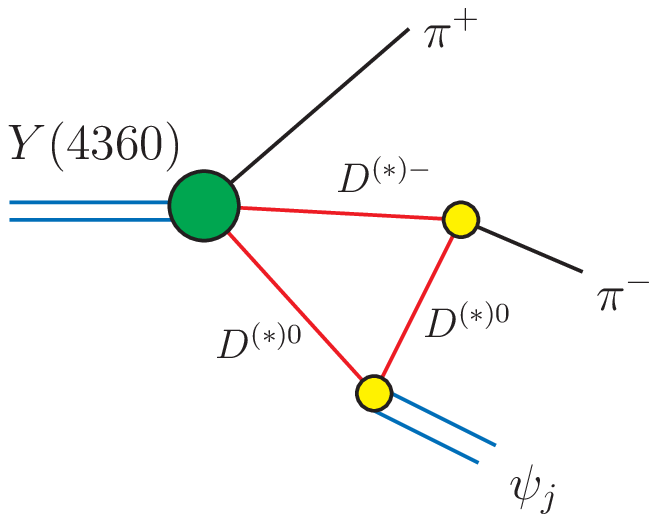} &
\includegraphics[width=0.22\textwidth]{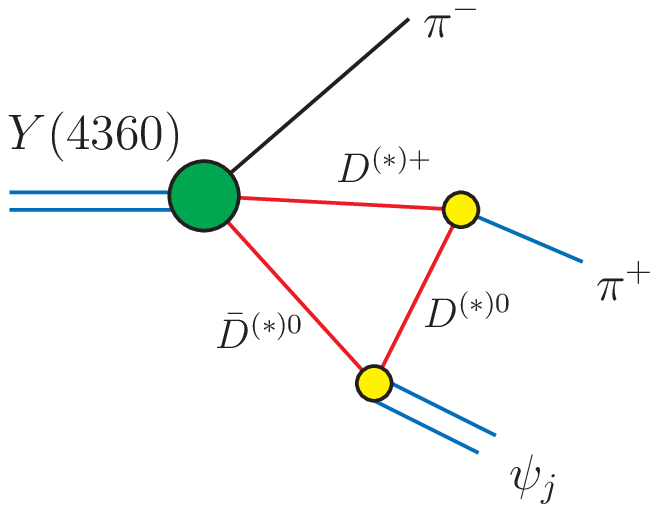} \\
$(a)$&$(b)$
\end{tabular}
\caption{(color online). The typical diagrams depicting $Y(4360)\to \psi_j \pi^+\pi^-$ $(\psi_{j}=J/\psi,\,\psi(2S),\,h_c(1P))$ by the ISPE mechanism.
}\label{Fig:typical}
\end{figure}

The involved hidden-charm dipion decays of $Y(4360)$ include:
\begin{eqnarray}
Y(4360)\to \pi^\pm [D^{(*)}\bar{D}^{(*)}]_{D^{(*)}}^\mp\to \left\{
 \begin{array} {l} J/\psi\pi^+\pi^-\\ \psi(2S) \pi^+\pi^-\\h_c(1P)\pi^+\pi^- \end{array} \right . , \label{eq1}
\end{eqnarray}
where $\pi^\pm$ is an initial single pion directly emitted from $Y(4360)$ decay. With the pion emission, $Y(4360)$ transits into intermediate
$D^{(*)}\bar{D}^{(*)}$. Then, via the exchanged $D^{(*)}$ meson, the intermediate
$D^{(*)}\bar{D}^{(*)}$ can dissolve into a charmonium plus a pion. In Eq. (\ref{eq1}), the subscript and superscript of
$[D^{(*)}\bar{D}^{(*)}]_{D^{(*)}}^\mp$ denote the total charge of the intermediate charmed mesons and the corresponding exchanged charmed meson.  These processes can be represented by the typical diagrams shown in Fig. \ref{Fig:typical}.

Here, the initial single pion emission plays an important role to these processes, i.e.,
because the pion carries off some of the $Q$ value, the intermediate $D^{(*)}$ and $\bar{D}^{(*)}$ with low momenta can easily interact with each other and transit into final states. These typical diagrams reflect main physical picture of the ISPE mechanism \cite{Chen:2011pv}.

First, we give a general expression of the differential decay width
for
$Y(4360)(p_0)\to \pi^+(p_3)[D^{(*)}(p_1) {\bar D}^{(*)}(p_2)]\to \pi^+(p_3)\pi^-(p_4)\psi_j(p_5)$,
\begin{eqnarray}
d\Gamma=\frac{1}{3}\frac{1}{(2\pi)^3}\frac{1}{32 M^3_{Y(4360)}}\overline{|\mathcal{M}|^2}d m_{\psi_j \pi^+}^2
dm^2_{\pi^+\pi^-},
\end{eqnarray}
where $m_{\psi_j \pi^+}^2=(p_3+p_5)^2$ and
$m^2_{\pi^+\pi^-}=(p_3+p_4)^2$. The overline denotes the average
over the polarizations of the $Y(4360)$ in the initial state and the
sum over the polarization of $\psi_j$ in the final state. Thus, in
the following our main task is to calculate $|\mathcal{M}|^2$
corresponding different processes.

As shown in Fig. \ref{Fig:typical}, $Y(4360)\to \psi_j \pi^+\pi^-$
via the ISPE mechanism can be described by these hadron-level
diagrams. Thus, we can write out the decay amplitude using the
effective Lagrangian approach. The interaction Lagrangians involved
in strong interaction vertexes, which are invariant under the isopin
$SU(2)$ symmetry, are,
\cite{Oh:2000qr,Casalbuoni:1996pg,Colangelo:2002mj}
\begin{eqnarray}
&&\mathcal{L}_{Y(4360) D^{(*)} D^{(*)} \pi}\nonumber\\&&=-ig_{Y DD
\pi} \varepsilon^{\mu \nu \alpha \beta} Y_{\mu} \partial_{\nu} D
\partial_{\alpha} \pi \partial_{\beta} \bar{D} + g_{Y D^\ast D \pi} {Y}^{\mu} (D \pi
\bar{D}^\ast_{\mu} + D^\ast_{\mu} \pi \bar{D}) \nonumber\\
&&\quad-ig_{Y D^\ast D^\ast \pi} \varepsilon^{\mu \nu \alpha
\beta} Y_{\mu} D^\ast_{\nu} \partial_{\alpha} \pi
\bar{D}^\ast_\beta -ih_{Y D^\ast D^\ast \pi} \varepsilon^{\mu \nu
\alpha \beta} \partial_{\mu} Y_{\nu} D^\ast_{\alpha} \pi
\bar{D}^\ast_{\beta},\nonumber
\end{eqnarray}
which depicts the initial state $Y(4360)$ interacts with $D^{(*)}
\bar{D}^{(*)} \pi$, and
\begin{eqnarray}
&&\mathcal{L}_{D^\ast D^{(\ast)} \pi} \nonumber\\&&= ig_{D^\ast D
\pi} (D^\ast_{\mu} \partial^\mu \pi \bar{D}-D \partial^\mu \pi
\bar{D}^\ast_{\mu})-g_{D^\ast D^\ast \pi} \varepsilon^{\mu \nu
\alpha \beta}
\partial_{\mu} D^\ast_{\nu} \pi \partial_{\alpha}
\bar{D}^\ast_{\beta},\nonumber
\end{eqnarray}
\begin{eqnarray}
&&\mathcal{L}_{\psi D^{(*)} D^{(*)}}\nonumber\\&&=ig_{\psi DD}
\psi_{\mu} (\partial^\mu D \bar{D}- D \partial^\mu \bar{D})-g_{\psi
D^\ast D} \varepsilon^{\mu \nu \alpha \beta}
\partial_{\mu} \psi_{\nu} (\partial_{\alpha} D^\ast_{\beta} \bar{D}
\nonumber\\&&\quad+ D \partial_{\alpha}
\bar{D}^\ast_{\beta})-ig_{\psi D^\ast D^\ast} \big\{ \psi^\mu
(\partial_{\mu} D^{\ast \nu} \bar{D}^\ast_{\nu} -D^{\ast \nu}
\partial_{\mu}
\bar{D}^\ast_{\nu}) \nonumber\\
&&\quad+ (\partial_{\mu} \psi_{\nu} D^{\ast \nu} -\psi_{\nu}
\partial_{\mu} D^{\ast \nu}) \bar{D}^{\ast \mu} + D^{\ast \mu}(\psi^\nu \partial_{\mu} \bar{D}^\ast_{\nu} -
\partial_{\mu} \psi^\nu \bar{D}^\ast_{\nu})\big\},\nonumber
\end{eqnarray}
\begin{eqnarray}
&&\mathcal{L}_{h_c D^{(*)} D^{(*)}}\nonumber\\&&= g_{h_c D^\ast D}
h_c^\mu (\bar{D}^\ast_{\mu} D + D^\ast_\mu \bar{D})+ ig_{h_c D^\ast
D^\ast} \varepsilon^{\mu \nu \alpha \beta}
\partial_{\mu} h_{c \nu} D^\ast_{\alpha} \bar{D}^\ast_{\beta}\nonumber
\end{eqnarray}
which are corresponding to the couplings of charmed meson pair with
pion or charmonium, where the charm meson iso-doublets can be
defined as $D^{(*)}=(D^{(*)0},D^{(*)+})$,
${\bar{D}^{(*)T}}=(\bar{D}^{(*)0},D^{(*)-})$ and
$\pi={\mbox{\boldmath $\tau$}}\cdot {\mbox{\boldmath $\pi$}}$
\cite{Oh:2000qr}.

In the heavy quark limit, the coupling constants among charmonia and
charmed mesons satisfy \cite{Casalbuoni:1996pg, Colangelo:2002mj},
\begin{eqnarray}
&&g_{\psi D D} = g_{\psi D^\ast D^\ast} {m_{D^\ast}}/{m_D} = g_{\psi
D^\ast D } m_{\psi} \sqrt{{m_D}/{m_D^\ast}} ={m_{\psi}}/{f_\psi},\label{couple1}
\nonumber\\\\
&&g_{h_c D^\ast D} =-g_{h_c D^\ast D^\ast} m_{h_c}
\sqrt{m_D/m_{D^\ast}}= -2g_1 \sqrt{m_{h_c} m_D m_{D^\ast}},\nonumber\\
\end{eqnarray}
where $f_{\psi}$ is the decay constant of a charmonium $\psi$. With
the center values of the leptonic partial decay widths, i.e.,
$\Gamma_{J/\psi\to e^+e^-}=5.55 \pm 0.14 \pm0.02\ \mathrm{keV}$ and
$\Gamma_{\psi(2S)\to e^+e^-}=2.35 \pm 0.04\ \mathrm{keV}$
\cite{Beringer:1900zz}, one can obtain $f_{J/\psi}=416\
\mathrm{MeV}$ and $f_{\psi(2S)}=298\ \mathrm{MeV} $. The gauge
coupling $g_1$ can be related to the decay constant of $\chi_{c0}$ through
$g_1=-\sqrt{m_{\chi_{c0}}/3}/f_{\chi_{c0}}$, where the $\chi_{c0}$ decay constant
$f_{\chi_{c0}}=510\ \mathrm{MeV}$ is estimated by the QCD sum rule
analysis \cite{Colangelo:2002mj}. Considering the chiral symmetry
and the heavy quark limit, the coupling constants among charmed mesons and pion
satisfy
\begin{eqnarray}
g_{D^\ast D^\ast \pi}=g_{D^\ast D \pi}/\sqrt{m_D m_{D^\ast}}
=2g/f_\pi,\label{couple3}
\end{eqnarray}
where  $f_\pi=132\ \mathrm{MeV}$ is the pion decay constant and
$g=0.59$ is extracted from the experimental measurement of the
partial decay width of $D^\ast \to D \pi$ \cite{Beringer:1900zz}.
The concrete values of the coupling constants in the effective
Lagrangians adopted in our paper are shown in Table
\ref{Tab:coupling}.

\begin{table}[htb]
\caption{The relevant coupling constants involved in the present
work. \label{Tab:coupling}}
\begin{tabular}{clclcl}
\toprule[1pt]%
Coupling & Value & Coupling & Value & Coupling & Value\\
\midrule[1pt] %
 $g_{J/\psi D D}$                & 7.44                        &
 $g_{J/\psi D^\ast D}$           & $2.49\ \mathrm{GeV}^{-1}$   &
 $g_{J/\psi D^\ast D^\ast }$     & 8.01                        \\
 $g_{\psi(2S) D D}$              & 12.39                       &
 $g_{\psi(2S) D^\ast D}$         & $3.49\ \mathrm{GeV}^{-1}$   &
 $g_{\psi(2S) D^\ast D^\ast }$   & 13.33                       \\
 $g_{h_c D^\ast D}$              & $15.21\ \mathrm{GeV}$       &
 $g_{h_c D^\ast D^\ast}$         & -4.77                       &\\
 $g_{D^\ast D \pi}$              & $17.31$                     &
 $g_{D^\ast D^\ast \pi}$         & $8.94\ \mathrm{GeV}^{-1}$  \\
\bottomrule[1pt] %
\end{tabular}
\end{table}

As an example of $Y(4360)\to J/\psi \pi^+\pi^-$ through intermediate
$D\bar{D}^*+H.c.$, we illustrate how to obtain the corresponding
decay amplitude. Although there are 12 diagrams for $Y(4360)\to
J/\psi \pi^+\pi^-$, we can find six independent diagrams since other
diagrams can be obtained by considering the $SU(2)$ symmetry. Thus,
the total decay amplitude reads as
\begin{eqnarray}
\mathcal{M}_{D\bar{D}^*+H.c.}^{J/\psi\pi^+\pi^-}=2 \sum_{\alpha=1,\cdots,6}\mathcal{A}^{(\alpha)}_{D\bar{D}^*+H.c.},
\end{eqnarray}
where the factor 2 is due to the isospin $SU(2)$ symmetry. Using the
effective Lagrangian approach, we get sub-amplitudes
$\mathcal{A}^{(1)}_{D\bar{D}^*+H.c.}$,
$\mathcal{A}^{(2)}_{D\bar{D}^*+H.c.}$  and
$\mathcal{A}^{(3)}_{D\bar{D}^*+H.c.}$, which correspond to the
dipion transitions between $Y(4360)$ and $J/\psi$ with an initial
single pion ($\pi^+$) emission , i.e.,
\begin{eqnarray}
&&\mathcal{A}_{D^\ast\bar{D}+h.c.}^{(1)}  = (i)^3 \int \frac{d^4
q}{(2\pi)^4} [g_{Y D^\ast D \pi} \epsilon_{\psi}^\mu] [ i
g_{D^\ast D^\ast \pi} (ip_4^\rho)]
\nonumber\\
&&\quad\times  [-ig_{J/\psi D^\ast D^\ast}
\epsilon_{J/\psi}^\nu(g_{\theta \phi} (-iq_\nu+ip_{2 \nu}) +  g_{\nu
\theta}  (ip_{5 \phi}+ iq_{\phi}) \nonumber\\ &&\quad -g_{\nu \phi}
(ip_{2\theta}+ip_{5\theta}) )] \frac{1}{
p_1^2-m_D^2}\frac{-g_{\mu}^\phi +p_{1 \mu}
p_{1}^{\phi}/m_{D^\ast}^2}{p_2^2 - m_{D^\ast}^2 }
\nonumber\\
&& \quad  \times \frac{-g_{\rho}^{ \theta} +q_\rho
q^\theta/m_{D^\ast}^2}{q^2-m_{D^\ast}^2}
\mathcal{F}^2(q^2,m_{D^*}^2),\label{ha1}
\end{eqnarray}
\begin{eqnarray}
&&\mathcal{A}_{D^\ast\bar{D}+h.c.}^{(2)} = (i)^3 \int \frac{d^4
q}{(2\pi)^4} [g_{Y D^\ast D \pi} \epsilon_{\psi}^\mu]
[ig_{D^\ast D \pi} (-ip_4^\rho)]
\nonumber\\
&&\quad \times [ig_{J/\psi DD} \epsilon_{J/\psi}^\nu
(ip_{2\nu}-iq_{\nu})] \frac{-g_{\mu \rho} +p_{1 \mu} p_{1
\rho}/m_{D^\ast}^2
}{p_1^2-m_{D^\ast}^2}\nonumber\\
&&\quad \times \frac{1}{p_2^2-m_D^2} \frac{1}{q^2-m_D^2}
\mathcal{F}^2(q^2,m_D^2),\label{ha2}
\end{eqnarray}
\begin{eqnarray}
&&\mathcal{A}_{D^\ast\bar{D}+h.c.}^{(3)} = (i)^3 \int \frac{d^4
q}{(2\pi)^4} [g_{Y D^\ast D \pi} \epsilon_{\psi}^\mu]
[-g_{D^\ast D^\ast \pi} \varepsilon^{\theta
\phi \delta \tau} (iq^\theta) \nonumber\\
&&\quad \times (-ip_1^\delta)] [-g_{J/\psi D^\ast D}
\varepsilon^{\rho \nu
\alpha \beta} (ip_{5 \rho}) \epsilon_{J/\psi \nu} (-iq_{\alpha})]\nonumber\\
&&\quad \times \frac{-g_{\mu \tau}+p_{1 \mu} p_{1
\tau}/m_{D^\ast}^2}{p_1^2-m_{D^\ast}^2} \frac{1}{p_2^2-m_D^2}
\nonumber\\&&\quad\times\frac{-g_{\beta \phi}+q_\beta
q_\phi/m_{D^\ast}^2}{q^2- m_{D^\ast}^2}
\mathcal{F}^2(q^2,m_{D^*}^2).\label{ha3}
\end{eqnarray}
Using Eqs. (\ref{ha1})-(\ref{ha3}), we can obtain the sub-amplitudes
$\mathcal{A}^{(4)}_{D\bar{D}^*+H.c.}$,
$\mathcal{A}^{(5)}_{D\bar{D}^*+H.c.}$  and
$\mathcal{A}^{(6)}_{D\bar{D}^*+H.c.}$ corresponding to the dipion
transitions between $Y(4360)$ and $J/\psi$ with an initial $\pi^-$
emission, where we need to interchange $p_3\rightleftharpoons p_4$
in Eqs. (\ref{ha1})-(\ref{ha3}).  Here,
$\mathcal{F}(q^2,m_E^2)=(\Lambda^2-m_E^2)/(q^2-m^2_E)$ denotes the
monopole form factor with the parameterized $\Lambda$, which is
taken as $\Lambda=m_E+\beta \Lambda_{QCD}$ with $\Lambda_{QCD}=220$
MeV and $m_E$ is the exchanged meson mass in the triangle hadron
loops. The parameter $\beta$ is expected to be of order unity
\cite{Cheng:2004ru}. In the present work we take $\beta =1$. In Ref.
\cite{Chen:2011xk}, we have numerically shown that the corresponding
lineshapes of the hidden-charm dipion decays of higher charmonia and
charmoniumlike state $Y(4260)$ are weakly dependent on the parameter
$\beta$. Considering the similarity between $Y(4260)$ and $Y(4360)$,
we can conclude that the discussed lineshape of the hidden-charm
dipion decays of $Y(4360)$ are also weakly dependent on the value
$\beta$.

In a similar way, we can also write out the decay amplitudes of
$Y(4360)\to J/\psi \pi^+\pi^-$ through intermediate $D^*\bar{D}^*$
and $D\bar{D}$, and $Y(4360)\to \psi(2S) \pi^+\pi^-$ and $Y(4360)\to
h_c(1P) \pi^+\pi^-$ through $D^{(*)}\bar{D}^{(*)}$ (see Ref.
\cite{Chen:2011xk} for more details).

We need to stress that in our work we calculate the individual
contributions of the intermediates, $D\bar{D}$, $D\bar{D}^*+H.c.$
and $D^*\bar{D}^*$, to the hidden-charm dipion decays of $Y(4360)$
concerned, and that we do not include the interference effects among
the decay amplitudes for different intermediate states. There are
two main reasons for this. The first one is that it is difficult to
obtain the concrete values of the coupling constants $g_{YD^{(\ast)}
D^{(\ast)} \pi}$ relevant to the internal structure of $Y(4360)$,
i.e., too many ambiguities to determine these constants. The second
is that the relative phases among different decay amplitudes cannot
be also constrained by present experimental data. In other words,
further experimental study of these novel phenomena discussed in
this work can make us carry out the fit to the experimental data
with our model. For example, after predicting the charged
charmoniumlike structures in the hidden-charm dipion decays of
$Y(4260)$ in Ref. \cite{Chen:2011xk},
having the new experimental data of $Z_c(3900)$ observed by BESIII
\cite{Ablikim:2013mio} we have succeeded in reproducing the
structure of $Z_c(3900)$ by including all the mechanisms
\cite{Chen:2013abc}, tree and other diagrams and relative phases,
among which the ISPE mechanism is the most dominant contribution to
the structure. Hence, following the same analysis, we can perform
the similar study on $Y(4360)$ if the detailed experimental data of
the hidden-charm dipion decays of $Y(4360)$ is obtained. Considering
the above reasons, in the present work we only concern only the
lineshapes caused by the ISPE mechanism with individual intermediate
states, where the maximum of the lineshape is normalized to 1 as
shown in the next section.

\section{Numerical results}\label{sec3}

\begin{figure}[htbp]
\begin{tabular}{c}
\includegraphics[width=0.45\textwidth]{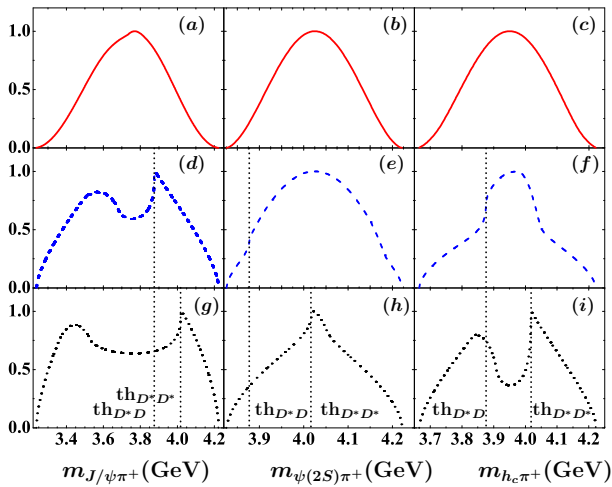}
\end{tabular}
\caption{(color online). The obtained  differential decay widths of
$Y(4360)\to \psi_j \pi^+\pi^-$ dependent on the invariant mass
spectrum $m_{\psi_j\pi^+}$. Here, the line shapes in the first, the
second and the third columns correspond to the cases taking
$\psi_j=J/\psi$, $\psi(2S)$ and $h_c(1P)$, respectively. The maximum
of these line shapes are normalized to 1.In addition, the vertical
dashed lines denote the $D\bar{D}^*$ and $D^*\bar{D}^*$ thresholds,
which are marked in these figures. The red solid, blue dashed and
black dotted curves are the results with $D\bar{D}$,
$D\bar{D}^*+H.c.$ and $D^*\bar{D}^*$ as the intermediate states,
respectively. \label{Fig:result}}
\end{figure}

With the formula derived in Sec. \ref{sec2}, we calculate the line
shapes of the differential decay width of $Y(4360)\to \psi_j
\pi^+\pi^-$ dependent on the $\psi_j \pi^+$ invariant mass spectrum,
which are shown in Fig. \ref{Fig:result}.

The results listed in Fig. \ref{Fig:result}  indicate:

(1) For $Y(4360)\to J/\psi\pi^+\pi^-$, there are two sharp peaks
existing in the $J/\psi\pi^+$ invariant mass spectrum. The one is
around the $D\bar{D}^*$ threshold while another one is close to the
$D^*\bar{D}^*$ threshold (See Fig. \ref{Fig:result} (d) and (g) ).
There are two broad structures, which are the corresponding
reflections of these two sharp structures.

(2) The line shapes of $\psi(2S) \pi^+$ invariant mass spectrum of
$Y(4360)\to \psi(2S)\pi^+\pi^-$ show different behavior from (1).
The intermediate $D\bar{D}^*+H.c.$ can not result in sharp peak
structures, instead there is a smooth curve (See Fig.
\ref{Fig:result} (e)). When including only $D^*\bar{D}^*$ as the
intermediate state, we find that a small peak exists although the
peak near the $D^*\bar{D}^*$ threshold and its reflection form one
peak as shown in Fig. \ref{Fig:result} (h).

(3) For the $Y(4360)\to h_c(1P)\pi^+\pi^-$ process, one irregular
peak exists in the $h_c(1P)\pi^+$ invariant mass spectrum if only
considering the intermediate $D\bar{D}^*+H.c.$ contribution (See
diagram Fig. \ref{Fig:result} (f)). The intermediate $D^*\bar{D}^*$
contribution can produce a sharp peak near the $D^*\bar{D}^*$
threshold (See Fig. \ref{Fig:result} (i)). Its reflection is a broad
structure, which is around the $D\bar{D}^*$ threshold. Comparing
with the sharp peak structure shown in Fig. \ref{Fig:result} (d),
the peak in Fig. \ref{Fig:result} (f) is more obvious.

In addition, our calculation also shows that the intermediate
$D\bar{D}$ contribution to $Y(4360)\to \psi_j \pi^+\pi^-$  cannot
result in any enhancement structure close to the $D\bar{D}$
threshold in the $\psi_j\pi^+$ invariant mass spectrum (See Fig.
\ref{Fig:result} (a)-(c)).

Since we only focus on the line shapes as shown in Fig.
\ref{Fig:result}, the uncertainties of the coupling constants cannot
result in the changes of our results. As for the $Y(4260)\to J/\psi
\pi^+\pi^-$ process via the intermediate state $D\bar{D}^*$ as an
example, a common factor $g_{YD^{(*)}D^{(*)}\pi} g /f_{\psi}$ can be
extracted from the total decay amplitude, which is due to the
constraint from Eqs. (\ref{couple1})-(\ref{couple3}). Thus, the
uncertainties of gauge coupling constants $g$ and $f_{\psi}$ cannot
change the corresponding line shapes in the $J/\psi\pi^\pm$
invariant mass spectrum.

Indeed, although in our calculation we use the relations of the
coupling constants that are obtained in the heavy quark limit, our
qualitative conclusion does not depend on these relations but on the
ISPE mechanism.

\section{summary}\label{sec4}

Heavy quarkonium physics is an intriguing and active research field
full of challenges and opportunities \cite{Brambilla:2010cs,
Brambilla:2004wf, Brambilla:2004jw}. A newly observed charged
charmoniumlike structure $Z_c(3900)$ was reported by BESIII
\cite{Ablikim:2013mio} and confirmed by Belle \cite{Liu:2013dau} in
the $Y(4260)\to J/\psi\pi^+\pi^-$ process. Before the BESIII's
observation, we have explored the hidden-charm dipion decays of
$Y(4260)$ and have predicted charged charmoniumlike structures by
the ISPE mechanism \cite{Chen:2011xk}.

Stimulated by the similarity between $Y(4360)$ and $Y(4260)$ and the
recent experimental observation of $Z_c(3900)$ by the BESIII and
Belle Collaborations, we study the hidden-charm dipion decays of
charmoniumlike state $Y(4360)$ in the present work. Our calculation
shows that there exist charged structures in the $J/\psi\pi^+$
invariant mass spectrum, which are near the $D\bar{D}^*$ and
$D^*\bar{D}^*$ thresholds (see Fig. \ref{Fig:result} (a) and (d)). A
peak structure around the $D^*\bar{D}^*$ threshold appears in the
$\psi(2S)\pi^+$ invariant mass spectrum (see Fig. \ref{Fig:result}
(e)). In addition, we can also find a sharp peak structure in the
$h_c(1P)\pi^+$ invariant mass spectrum near the $D^*\bar{D}^*$
threshold (see Fig. \ref{Fig:result} (f)).

Experimental search for these charged charmoniumlike structures
predicted near the $D\bar{D}^*$ and $D^*\bar{D}^*$ thresholds is an
interesting research topic, which will be helpful to further test
our predictions presented in this work. What is more important is
that the ISPE mechanism existing in the hidden-charm/hidden-bottom
dipion decays of higher charmonium/bottomonium can be tested again.
BESIII, Belle, and forthcoming BelleII will be a good platform to
carry out the experimental study on these hidden-charm dipion decays
of charmoniumlike state $Y(4360)$. We would like to have more
experimental progresses in this field.

\vfil
\section*{Acknowledgement}

This project is supported by the National Natural
Science Foundation of China under Grants No. 11222547, No. 11175073,
No. 11005129 and No. 11035006, the Ministry of Education of China
(FANEDD under Grant No. 200924, SRFDP under Grant No. 20120211110002, the Fundamental Research Funds for the Central
Universities), the Fok Ying Tung Education Foundation (Grant No. 131006),
and the West Doctoral Project of Chinese Academy of Sciences.

\end{document}